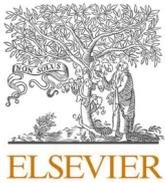
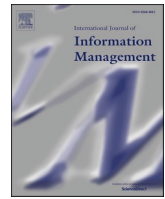

Research Article

# Adaptive cognitive fit: Artificial intelligence augmented management of information facets and representations

Jim Samuel [a],[*], Rajiv Kashyap [b], Yana Samuel [c], Alexander Pelaez [d]

[a] *Rutgers University, NJ, United States*
[b] *William Paterson University, NJ, United States*
[c] *Northeastern University, MA, United States*
[d] *Hofstra University, NY, United States*

ABSTRACT

Explosive growth in big data technologies and artificial intelligence (AI) applications have led to increasing pervasiveness of information facets and a rapidly growing array of information representations. Information facets, such as equivocality and veracity, can dominate and significantly influence human perceptions of information and consequently affect human performance. Extant research in cognitive fit, which preceded the big data and AI era, focused on the effects of aligning information representation and task on performance, without sufficient consideration to information facets and attendant cognitive challenges. Therefore, there is a compelling need to understand the interplay of these dominant information facets with information representations and tasks, and their influence on human performance. We suggest that artificially intelligent technologies that can adapt information representations to overcome cognitive limitations are necessary for these complex information environments. To this end, we propose and test a novel "Adaptive Cognitive Fit" (ACF) framework that explains the influence of information facets and AI-augmented information representations on human performance. We draw on information processing theory and cognitive dissonance theory to advance the ACF framework and a set of propositions. We empirically validate the ACF propositions with an economic experiment that demonstrates the influence of information facets, and a machine learning simulation that establishes the viability of using AI to improve human performance.

## 1. Introduction

*'We have begun a conversation with machines that will last for the rest of our lives - that will also be remembered by those machines long after our own fragile memories have failed us.'* William Ammerman 2019

Big data and artificial intelligence are changing the essence and form of our interactions with information and with information processing machines. No longer are human emotions confined to interpersonal relationships – instead, we find ourselves displaying genuine feelings and sentiment when confronted by intelligent agents and systems (Law, Chita-Tegmark, & Scheutz, 2021). Further, we have witnessed significant increases in information entropy due to social media data of questionable veracity and deliberate attempts to provide misinformation (Moravec, Minas, & Dennis, 2018). As a result, decision makers are confronted by information embedded with greater equivocality, veracity, and density. These developments portend *fundamental changes* in the nature of information that necessitate attention to the ways in which information is received, conceived, interpreted and acted upon. To this end, we conceptualize and develop the Adaptive Cognitive Fit (ACF) framework, to improve our understanding of how humans can and should leverage artificially intelligent technologies to make decisions in the emerging complex information ecosystems. "*Adaptive*" in this context refers to the use of artificially intelligent methods to dynamically evolve the representation of information to improve their functional affordances and symbolic value (DeSanctis & Poole, 1994; Markus & Silver, 2008). For instance, AI agents can be deployed to augment human intelligence by mining data (e.g., video, voice, text) in real-time to improve employee interactions with an organization's stakeholders (Davenport, Guha, Grewal, & Bressgott, 2020). Further, AI morphed representations of information artifacts (e.g., avatars, colors, images, personal intelligent agents) can enhance user experiences by adapting environments through attention to culture, semiotics, and anthropomorphism (Moussawi, Marios, and Benbunan-Fich, 2021; Nantheera and Bull, 2022). These affordances represent a small fraction of the vast range of applications and communicative potential of information artifacts in the context of adaptive and intelligent systems.

The present age of big data and ubiquitous connectivity is characterized by explosive demand for business analytics and AI systems that enhance *decision making* and set high performance expectations for *decision makers* (Kushwaha, Kar, & Dwivedi, 2021; Dwivedi et al., 2019). The increasing use of data visualization, machine learning, programming languages, adaptive methods, optimization, and automation have catalyzed these trends (e.g., Cavalcante, Frazzon, Forcellini, & Ivanov, 2019; Kar & Dwivedi, 2020). This is evidenced by the rapid growth of the big data and business analytics market, which is projected to reach $274.3 billion by 2022 (Mlitz, 2021). However, these new information environments have created significant ambiguity about the alignment of information representations with decision needs to optimize human performance. These representations include varying combinations of static and dynamic displays, tabular summaries, and interfaces characterized by increased complexity and diversity of information representations. While there is a fair amount of research highlighting the need for cognitive alignment of technologies to tasks, most studies were conducted in the pre-big-data era of information systems (e.g., Bacic & Henry, 2018; Vessey, 1991; Vessey & Galletta, 1991). The need to develop artificially intelligent and adaptive frameworks to align information representations with tasks in the new information ecosystem of big data, AI, and business analytics remains unmet (Dwivedi et al., 2020).

CFT research guided decision makers to "*choose the data display that is appropriate to the type of task…*" and urged organizations to automate data representation systems to "*provide data in an optimal format*" aligned with the problem type (Vessey & Galletta, 1991). This alignment helped to maximize the accuracy and efficiency of mental representation of the task or problem, leading to superior performance. However, the advent of big data, business analytics, and AI has created new decision-making challenges due to greater informational complexity. These problems are compounded by streams in current IS research which tend to generate insights through less generalizable data centric approaches, while ignoring the need for theory building relevant to management disciplines (Kar & Dwivedi, 2020).

Along with information related challenges, the range and types of information representations have changed significantly. Therefore, there is an urgent need to improve our understanding of how big data amplified information facets such as equivocality and veracity influence performance. Drawing on information processing theory and cognitive dissonance theory, we expand the concept of cognitive fit to address the needs of the big data and AI ecosystem.

Therefore, in the context of big data and AI ecosystems, the research questions we pose are:

1. How do information facets, such as equivocality, influence performance?
2. How can AI help manage cognitive fit in complex information environments?

We present the Adaptive Cognitive Fit (ACF) framework as a novel articulation of information facets, information representations and artificial intelligence for maximizing performance (Figs. 1, 2 and 4). ACF[1] addresses a critical need by presenting and validating the role of information facets in decision making and corresponding performance. We posit that cognitive fit needs to be adaptive. In this study, "*adaptive*" reflects the use of AI methods, technologies and informatics to evolve and dynamically align information representations with variations in information facets, tasks, and human intelligence, to improve performance.

---

[1] See Table 1 for a list of abbreviations used to ease exposition.

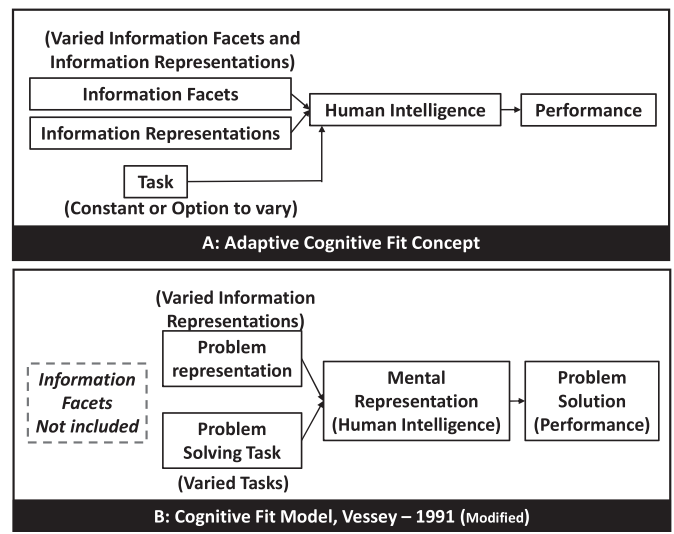

**Fig. 1.** A,B ACF Concept – Influence of Information Facets.

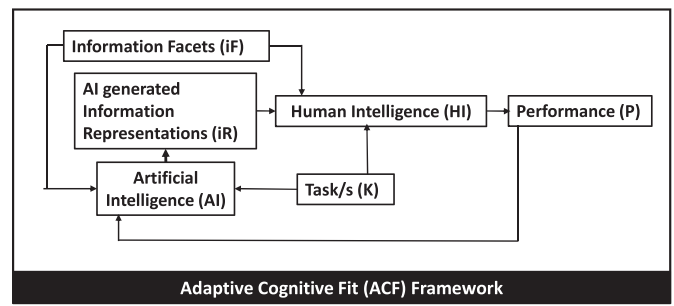

**Fig. 2.** ACF Framework.

In the sections that follow a review of extant research on cognitive fit, we present ACF in four parts, and then conclude. First, we elaborate the theoretical basis underlying the ACF framework and generate four ACF propositions. Next, we develop hypotheses to demonstrate the influence of information facets on performance. In the third part, we present an experiment and discuss its results. In the fourth part, we deploy an unsupervised machine learning simulation to illustrate an AI application of ACF. We conclude with a discussion that highlights the implications and limitations of ACF and identifies future research opportunities.

## 2. Big data and cognitive challenges

The big data phenomenon has immersed mankind into the paradigmatic "*dark side of information*" (Grover, Lim, & Ayyagari, 2006). Information has grown in power and dimensionality, and its transformative potential and evolution need to be factored into all research that engages information artifacts (Chen, Chiang and Storey, 2012; Kar & Dwivedi, 2020; McKinney & Yoos, 2010; Rai, Constantinides, & Sarker, 2019). Extant CFT research has not sufficiently investigated the "*dark side of information*" (Grover et al., 2006) and progressive changes in the meanings, dynamics, and use of information (McKinney & Yoos, 2010). Therefore, there is a critical need to develop models that relate information facets such as equivocality and veracity to performance based on cognitive fit (Grover et al., 2006; Samuel, 2016). This lacuna in prior research warrants immediate attention, as almost every meaningful task, objective, problem, or decision in the future will involve human intelligence (HI) perceptions (Rai et al., 2019) that are conditioned by information facets. The extremity and pervasiveness of information facets have created additional challenges for finding cognitive fit solutions in information-rich environments (Hills, 2019). Fortunately, significant advances in AI methods and technologies are now available to adapt information representations and improve performance. Therefore, we aim to fill the gap in cognitive fit research by developing a theoretical framework that explains the influence of information facets and adaptive AI augmentation on performance.

### 2.1. The rise of information facets

"**Information facets**", which can be viewed as the big data and AI era kernels of information that influence cognition, need to be a part of any discourse on artificially intelligent and adaptive cognitive support for human intelligence. Information facets such as equivocality, veracity, and overload can be viewed as quintessential properties of information, which when sufficiently dominant, have the potential to influence perception, and hence performance. Information facets can be pervasive, persistent, variable, and consequential. However, their influence on human cognition is contingent on the scope of their presence and intensity, elements of the information environment that stimulate perception, decision support resources, and individual expertise and knowledge. Weak levels of information facets may influence individuals differently. Therefore, when an information facet is sufficiently dominant, its influence is fundamental and will engender a common experience. This is evidenced by extant research that has validated the statistical significance of equivocality and also by the experiment in this study (Brecher & Hantula, 2005; Hantula & DeNicolis, 1999; Dennis & Kinney, 1998). Therefore, we define "information facet" as the latent but characteristic essence of an information artifact, which when sufficiently dominant, directly impacts human perception of meaning, content, and substance, and consequently influences performance (Samuel, 2016, Grover et al., 2006).

For example, consider a situation in which a decision maker utilizes information that is bounded and yet intrinsically of high uncertainty. *All other things being equal*, and task-agnostic, performance can only be managed to a limited extent by changes to information representation. However, transforming the available information to low uncertainty can improve performance as the decision maker would be less prone to cognitive selection biases (Hills, 2019).

### 2.2. Information representations and emerging challenges

"**Information representations**", in contrast, refers to the structure of tangible information artifacts such as graphs and tables (Vessey, 1991). In the current context, information representations include the naturally sensible static and dynamic human intelligence tangible expressions of information, such as text, data visualizations, signs-symbols, data tables and summaries, data models, audio, video, and braille, and variations of these. Accordingly, we consider information representations as possessing properties that are 'created' within the context of the decision-making environments (McKinney & Yoos, 2010). From this perspective, information representations can possess subjective and temporal bearings, and importantly, each information representation can reflect information scenarios with varying levels of latent information facets. Big data phenomena and the science of data analytics have led to a massive growth in the number, diversity and forms of information representations such as user-generated social media content, machine-generated data from IOT applications and complex 3D visualizations (Kar & Dwivedi, 2020; Walden, Cogo, Lucus, Moradiabadi, & Safi, 2018). Manually mapping such a diverse and expanding set of information representations using extant cognitive fit models would be extremely inefficient. New technological solutions are needed to augment human intelligence and facilitate cognitive fit.

### 2.3. Past efforts towards cognitive fit

Past efforts at framing cognitive fit in the pre-big-data era indicated that decision makers could use cognitive fit theory (CFT) to "*choose the data display that is appropriate to the type of task…*" and alternatively that organizations could manage data representation systems so that "*they provide data in an optimal format*" that is aligned with the problem type (Vessey & Galletta,1991; Welles & Xu, 2018). Similarly, much of the early work on CFT in information systems and decision sciences revolved around, was connected to, or was based upon the contrasts between graphics and numbers or text in tables. The primary emphasis was on the alignment of information representation (in the form of tables or graphics) with the task such that the aligned information representation maximized the accuracy and efficiency of mental representation of the task or problem, leading to superior performance (Fig. 1-B). However, with the advent of big data, analytics and artificial intelligence (AI), the informational ecosystem surrounding tasks and decision-making challenges have expanded in size, speed of information flow, fragmentation, options and complexity. Therefore, there is a need to expand the traditional cognitive fit perspective to improve our understanding of how complex information facets such as information equivocality and veracity influence performance.

### 2.4. Theoretical background: from cognitive fit to adaptive cognitive fit

In this section, we discuss three prominent streams of theories which contribute to a better understanding of ACF: cognitive fit, information processing and cognitive dissonance theories. Cognition, as a broad capability of HI, goes well beyond being limited to the concept of identification or recognition and deals with how humans "*encode, structure, store, retrieve, use*" information (Lutz & Huitt, 2003; Neisser, 1967). In the context of complex big data environments, cognition needs to be addressed from processing and dissonance management perspectives to frame fit and performance. Recent research has indicated that AI can be used to reduce cognitive dissonance (Lee & Joshi, 2020). We also draw attention to recent IS research which has called for the use of AI, machine learning and associated methods to develop IS theories, and artificially intelligent systems and solutions (MISQ Special Issue, 2021).

### 2.5. Cognitive fit – the need for more

CFT research covering external and internal dimensions of information representation, problem solving abilities, task characteristics and complexity of tasks, has been widely discussed, developed, and supported (Chandra & Krovi, 1999; Shaft & Vessey, 2006; Cardinaels, 2008; Baker, Jones, & Burkman, 2009). Past IS research posited that performance improves if there is a fit between the task and the technology used to address the task (Goodhue & Thompson, 1995; Zigurs & Buckland, 1998). CFT presents an underlying theme of cognitive effort applied to spatial and symbolic tasks, with the premise that a lack of alignment and 'fit' between task and information representation increased cognitive effort in mental representation and therefore adversely affected performance. Conversely, the presence of alignment and 'fit' between task and information representation decreased cognitive effort in mental representation and therefore improved decision performance. Extant studies placed emphasis on accuracy and speed of decision output as dependent measures for cognitive fit (Vessey, 1991). An alignment between the task and representation supported the formation of mental representations that led to higher accuracy and speed, than if such an alignment or 'fit' were absent, or lower.

Fit has also been operationalized alternately as moderation, mediation, matching, holistic and covariation in previous IS research (Hoehle & Huff, 2012). Early developments of CFT emphasized the alignment between task and information representation, contrasting performance based on tables and graphs (Vessey, 1991; Vessey & Galletta, 1991), and also evaluated the alignment of tool with task (Agarwal et al., 1996). Subsequent studies extended and vetted Vessey's (1991) cognitive fit model. Hong, Thong, and Tam (2004) compared 'list' with 'matrix' for online shopping and Dennis and Carte (1998) applied CFT to study how subjects using maps (geographic information systems) performed under different task conditions. Later, the CFT research stream widened to examine domain knowledge influence (Khatri, Vessey, Ramesh, Clay, & Park, 2006), mobile representation of online content (Adipat et al., 2011), self-reported cognitive effort (Bacic & Henry, 2018) and 3D visualizations (Walden et al., 2018). Fig. 1-B is a generalization of traditional cognitive fit models (Vessey, 1991). From a simplified perspective, CFT logic states that, everything else being constant, optimal or best possible performance is accomplished by matching tasks with appropriate representations. Past research has focused almost exclusively on the nature of the task and type of information representations, which is schematically illustrated in Fig. 1-B. In contrast, we analyze if performance for the same task could vary with the same information representation, based on differences in an information facet (Fig. 1-A). Consistent support for CFT has often been challenged by qualifications, and identification of inconsistencies such as a need for hybrid, multiple and combined information representations, dimensions of graphical representation, task complexity challenges, need to factor cognitive effort, task specificity and individual style, and subjectivity to sub-types of information representations (Frownfelter-Lohrke, 1998; Kelton, Pennington, & Tuttle, 2010; Dennis & Carte, 1998; Speier, 2006; Bacic & Henry, 2018; Welles & Xu, 2018). ACF provides a basis to overcome several limitations of the traditional approach to cognitive fit. **First,** CFT is not designed to readily accommodate the variable nature of information facets, such as equivocality, veracity, and overload. ACF specifically aims to address the challenges that result therefrom. **Second,** CFT, as a pre-big data and AI era theory, could not have factored in the sophistication of present-day information representations, data visualizations, data tabulations, data varieties and data outputs (Kar & Dwivedi, 2020). For example, big data analytics driven visualizations or "*graphs*" often contain "discrete data values" (assumed under CFT as a property of tables), and similarly directional summary tables often possess less detail than corresponding visualizations, though "*derived from equivalent data*" (Vessey, 1991). CFT arguments are hence based on an outdated notion of 'purity' of critical constructs such as 'graph' and 'table'. **Third,** CFT is task-centric and limited by the consideration of variance in task specifics. Consequently, all consideration of information is task-bound. ACF overcomes this and the previous limitation by proposing a task-agnostic (task is constant, with an option to be varied) framework that prioritizes the influence of information facets on human performance (Fig. 1-A). **Fourth,** in the context of the big data ecosystem, CFT was not developed with an intent to support AI applications. For instance, Dwivedi et al. (2019) identify several social, economic, data, technology, political, legal, and ethical challenges to the successful development of AI applications. They express reservations due to the scope and quantum of work needed to assign meanings to deep learning techniques, the difficulty of including moral and ethical perspectives into AI applications in domains as varied as manufacturing and healthcare and the need for data integrity and transparency. ACF provides a starting point to address these challenges as it explains the interplay of human intelligence and information facets, which can be used to build AI systems that augment human capabilities and improve decision making.

*2.6. Information facets and cognitive limits – information processing theory*

Unaided HI is significantly limited in its capabilities to process information. Scholars have long recognized these limitations based on human inability to cope with the cognitive demands of acquiring, processing, and making sense of all available information (e.g, Hyman, 1953; Simon, 1957; Stalnaker, 1991). Human information processing has been defined as "*how information is modified so that it eventually has its observed influence*" (Massaro & Cowan, 1993). Information processing theory highlights how HI is bound by "*severe limitations on the amount of information that we are able to receive, process and remember*" (Miller, 1956). Social and organizational information processing models address how human intelligences interact, use relationships and leverage technology to make sense of information (Chidambaram, 1996; Gattiker & Goodhue, 2005; Galbraith, 1973, 1974). IS research has used information processing theory for studying the alignment of organizations and their technology architecture (Anandarajan & Arinze, 1998), healthcare information management (Bolon, 1998), analysis of vendor performance (Oshri, Dibbern, Kotlarsky, & Krancher, 2019) information processing in electronic networks (Meservy, Fadel, Kirwan, & Meservy, 2019), impression development (Cummings & Dennis, 2018) and information, feature and social overload (Fu, Li, Liu, Pirkkalainen, & Salo, 2020). However, there is a significant gap in extending information processing theory to explain cognitive fit in the context of big data, analytics and AI due to limitations of judgment and memory, the subjectivity of interim sensemaking (recoding) and the incalculable number of options to manipulate information and its effects (Hills, 2019).

*2.7. Information facets, dual information processing, and cognitive dissonance*

HI uses numerous mechanisms to manage its own information processing limitations. Individuals use a dual information processing strategy (Maheswaran & Chaiken 1991; Moskowitz et al., 2005), wherein they process information with less or more cognitive effort, depending upon the nature of information and environmental factors (e. g., time constraints, and complexity). The default strategy is to minimize cognitive effort by deploying sensemaking tools such as heuristics, schemas, stereotypes, or expectations to generate meaning. For instance, an upward trending line in a figure is congruent with an expectation of increase or growth. This 'least effort' rule prevails when information entropy is low, and individuals do not feel restricted by environmental factors (Petty & Cacioppo, 1986). However, changes in information facets or environmental conditions can trigger cognitive dissonance (Festinger, 1962). Cognitive dissonance has been described as the experience of human intelligences supplied with two or more elements of incompatible information which cannot be easily concluded upon (Moravec, 2018; Festinger, 1962). As an experience, cognitive dissonance can be emotional, psychological, visual or physical, often leading to increased levels of emotional and mental stress. The general tendency for HI is to try and reduce stress by either resolving the inconsistencies or by avoiding them. Assuming sufficient, incentive, HI will attempt to resolve the informational inconsistency. Based on the heuristic systematic model, which posits a dual information processing framework, there can be two ways to process information: heuristic-environmental and systematic (Li, Tan, Wei, & Wang, 2017; Tam & Ho, 2005).

Individuals attempt to resolve dissonance through deeper and more systematic processing. Such efforts may include cognitive efforts to reassess prior beliefs and expectations, and carefully deliberate, analyze, and validate the veracity of the information. Individuals may also attempt to relieve dissonance by adding consonant cognitions, subtracting dissonant cognitions, increasing the importance of consonant cognitions, or decreasing the importance of dissonant cognitions (Harmon-Jones, Harmon-Jones, & Levy, 2015). In other words, human intelligence reconciles differences in cognitions by recoding information to reduce dissonance. It follows that any augmentation that improves cognitive fit by adapting information representations will improve performance. While IS research has studied how human intelligences interact, relate, and leverage technology to make sense of information (Chidambaram, 1996; Gattiker & Goodhue, 2005; Galbraith, 1973, 1974), little attention has been given to the potential cybernetic enhancements (Wiener, 1948) that can be generated by augmenting HI with artificially intelligent technologies to overcome dissonance. Next, we describe the process by which AI augmentation assists human intelligence to reduce cognitive dissonance.

## 2.8. The ACF framework

Based on the discussion above, we describe the influence of information facets on HI to develop and present the ACF framework in this section (Fig. 2). Contextualized to current and emerging big data and AI ecosystems, ACF provides an expanded framework that includes and elaborates upon the interaction between prominent information facets, information representations, AI, and HI, to achieve the best possible or optimal performance. When one or more information facets dominate a task environment, it becomes harder for HI to form meanings and make effective mental connections with information artifacts. Individuals may incorrectly weigh performance criteria, exhibit risk aversion, or avoid making decisions (Montesano, 2019). AI technologies can be used to improve performance by generating information representations that reduce bias, lower cognitive loads, and relieve emotional stress. Improved performance is achieved through cognitive fit between information representation (iR) and human intelligence (HI) for a given task (K). ACF leans on this traditional conceptualization of cognitive fit. However, unlike traditional cognitive fit, ACF acknowledges information facets and posits that when an associated information facet (iF) changes significantly, optimal performance for the same original task (K) can be achieved by a new cognitive fit between an alternate information representation (alternate iR) and HI. This leads us to the first ACF proposition:

**P1**. A dominant information facet (iF) will influence performance (P) for a given task (K).

In the ACF framework, AI helps generate better information representations through iterative learning and adaptation processes to accommodate dynamic changes in information facets. Cognitive fit is induced by AI managed (error corrective) feedback and (error preventive) feedforward processes (Basso & Belardinelli, 2006). The AI component in ACF utilizes the feedback loop from performance (P) to identify the magnitude and characteristics of performance errors from variations in information facet levels. Subsequently, the feedforward loop anticipates and predicts the appropriate information representation category for achieving optimal performance for the same original task. Thus, based upon the prior-states interactions observed between information facets and information representations, AI generates improved choices for information representations, and learns to support optimal outcomes. ACF infuses cognitive fit logic with the control dynamics of cybernetics processes to develop artificially augmented solutions that optimize performance (Fig. 2).

## 2.9. ACF proposition 2

We have emphasized the critical need to recognize the influence of information facets upon human intelligence, in the context of big data, analytics and AI. Human intelligence possesses the capability to form internal mental representations based on external information facets and external information representations, for a given task or set of tasks. For ACF, we consider moderate to complex tasks, as HI's past knowledge and experience may subdue the effects of information facets and information representations for simple tasks. Under ACF, we posit that levels of iF and iR will jointly influence performance (P), and more pertinently, that an iF can potentially interact with iR such that for the same task K, a different iR may need to be used to achieve optimal performance, given a sufficient change in iF. This leads us to the second ACF proposition:

**P2**. Information facets (iF) will impact performance (P) through interaction effects with associated information representations (iR).

Though ACF propositionally accommodates multiple information facets, we identified equivocality as a powerful and illustrative iF with a high potential for interaction with iR and therefore focused specifically upon equivocality as the iF of interest in our economic experiment presented later.

## 2.10. Artificial intelligence

In this section, we explain the artificial intelligence (AI) dimension of ACF for managing the growing challenges posed by the vast and growing arrays of information representations and pervasive information facets. AI techniques have been used to classify human performance and to model human information behavior (Abubakar, 2019; Samuel, 2017). Given the challenges posed by the expanding variations of information facets, the increasing multiplicity of information representations, and the limitations of human intelligence, it is impossible to manually research, develop and continually update information representations and task-specific customized models of cognitive fit. Therefore, we employ machine learning, a prominent part of AI, to highlight the critical benefit of interactively and iteratively adapting and aligning information representations with varying information facets to drive optimal performance.

Recent cognitive research has recognized the need for "*…the development of smart interfaces that can prioritize appropriate data representations depending on the problem-solving task someone aims to complete*" (Welles & Xu, 2018). It has become increasingly evident that we need AI solutions to manage cognitive fit. Though AI has been an important discipline over the past few decades, it is only with the advent of big data and advanced computing capabilities that AI has become a critical part of the information technology expanse (Rai et al., 2019; Constantinides, Henfridsson, & Parker, 2018). AI has been defined from a variety of perspectives and is generally understood to be a collection of technologies, mathematical and statistical models, and vast arrays of diverse types of data, supported by an increasing number of applications that seek to model or exceed human intelligence functions. Within the IS discipline, AI has been parsimoniously and effectively defined as being "*the ability of a machine to perform cognitive functions that we associate with human minds, such as perceiving, reasoning, learning, interacting with the environment, problem solving, decision-making, and even demonstrating creativity*" (Rai et al., 2019). Additionally, the direction of recent IS research affirms the need and the viability of intelligent agents and adaptive systems which can "*can now assume responsibility for tasks with ambiguous requirements and for seeking optimal outcomes under uncertainty*" (Baird & Maruping, 2021). Artificial intelligences afford humankind the ability to stretch beyond previously forbidden limits and conceive new products, services and systems that can vastly impact and improve the quality of human life, and the environment we live in (Samuel, 2021). These are possible, in part because of breakthroughs in statistical computing over the past two decades, being applied in the context of big data as "machine learning".

## 2.11. Machine learning (ML)

Machine learning (ML), enabled by advances in statistical computing, refers to the computational processes by which machines can self-learn through organized iterations, with limited or without human intervention. ML is a prominent aspect of AI, and akin to nature-inspired swarm intelligence and generative adversarial networks methods, ML is treated as a subset of AI (Chakraborty & Kar, 2017; Aggarwal, Mittal, & Battineni, 2021). ML falls into three broad categories, including supervised, unsupervised, and reinforcement learning. These are augmented by an increasing number of powerful models that take advantage of newer forms of optimization, deep learning, neural architecture, recurrent and convolutional networks, and adversarial training. Unsupervised and supervised machine learning methods have been leveraged for clustering, classification and prediction in IS research (Ben-Assuli & Padman, 2020) and connecting systems, data and humans to improve healthcare (Bardhan, Chen, & Karahanna, 2020). Machine learning has also been used to analyze visual and textual social media data (Shin et al., 2020) and improve the performance of risk profiling processes through multitask learning (Lin, Chen, Brown, Li, & Yang, 2017). In this study we apply machine learning methods to model, simulate and classify human performance (Bardhan et al., 2020). We utilize an unsupervised machine learning model to test and validate the viability of an AI-augmented ACF framework.

*2.12. ACF propositions 3 and 4*

*AI can **learn** by itself.* We leverage one of the most critical aspects of AI, namely the ability of AI technologies to learn to perform intelligent activities, such as learning to play complex board games like chess and supersede HI capabilities for the same. The vast arrays of information representations from the big data ecosystem are too complex to be comprehensively addressed by research in traditional CFT fashion. Comparing, contrasting, and aligning each information representation to a near-infinite range of practitioner tasks is highly impractical. Amidst dynamic variance in information facets, the human identified information representation approach will become restrictive and non-optimal. This is demonstrated in the experimental study presented in this paper, where a change in the level of equivocality (iF) changes the best information representation (iR) option. However, it is possible to use AI techniques, including adaptive algorithms, to train applications to self-learn as more relevant data becomes available, leading to increased algorithmic interaction and user experiences (Shin, Zhong, & Biocca, 2020). Therefore, practical cognitive fit implementation in an organizational or commercial context will require artificially intelligent information systems and design science solutions that can **learn to** match the *scale* of information representations assortments to the *complexity and range* of evolving information facets, while accounting for the *subjectivity* of human intelligence (HI).

We use the term human intelligence in the context of ACF to specifically refer to the information processing, memory, perception, and cognitive properties of HI, including the capability to form "internal representations". Extant research has demonstrated that diverse information representations can have varying impacts on users' perceptions. This is evident in the use of domain-specific symbols, wherein human intelligences with training or past experience in utilizing mathematical symbols perform math related tasks faster and better than those without prior exposure to those symbols, or affinity for math. Past research has shown that users are sensitive, for example, to the dimensions of graphic information representations (Kumar & Benbasat, 2004), based on experience, expertise, cognitive bias and cultural sensitivity to information interfaces (Vessey, 1991; Reinecke & Bernstein, 2013). Decision and performance variables are representative of HI capabilities, and can be used as input for AI technologies. Hence, it is important for human intelligence capabilities to be integrated into any cognitive fit solution that claims practical relevance. Accordingly, the ACF articulation is based upon progressive, interactive, and iterative machine learning based adaptations to HI capabilities. Such artificially intelligent adaptations build on interactions between information representations and information facets, with individual or collective capabilities (as in teams), and "learn" to select appropriate information representations. This brings us to the third ACF proposition:

**P3**. *Artificially intelligent cognitive fit solutions learn to adapt to variations of information facets (iF), information representations (iR), tasks (K) and prior performance to generate progressively improved information representations.*

*AI can optimize* traditionally difficult to model scenario variables. AI methods such as deep learning and adversarial strategies have successfully led to optimal results, increasingly superior to traditional optimization modeling approaches. ACF emphasizes the need to use AI to identify optimal fit by accommodating variance in information facets, information representations, and tasks. Although novel in the context of using AI for cognitive fit, this approach to accommodating individual HI capabilities has been recommended for *"match-making …between user abilities and application"* (Recker, 2010; Welles & Xu, 2018). Exhaustive manual mapping of facets, tasks, representations, and capabilities can be too laborious using traditional methods, leading to non-optimal performance. However, it is possible to use AI techniques, including neural networks, deep learning, adversarial strategies and reinforcement learning, to train programs to move towards optimality with additional iterations and data.

This can lead to optimal cognitive fit solutions based on dynamically adaptive information representations aligned with information facets and task(s), which surpass manual efforts and even the automation scope of traditional information systems. For example, a broad range of machine and deep learning techniques, such as modular neural networks, can be organized to factor in multiple influences and interactions to dynamically learn and improve output to move towards optimality as more raw data and performance feedback data become available. Extant research has supported the use of AI to adaptively support cognitive and perception functions of HI and also demonstrated the behavioral influence of information facets on performance (Wan et.al., 2020; Wang et. al., 2020; Samuel, 2017). Therefore, we posit that AI generated (or classified or recommended) information representations can improve performance, and consequently, continued AI augmentation can lead to increasingly sustainable optimal performance. AI augmentation in ACF can thus help move towards optimal performance by continually improving the generation of information representations, using data from past scenarios, associated information facets, tasks, and past performance (Figs. 2 and 6).

This leads us to our final ACF proposition which posits that in an ever-expanding big data, analytics and advanced technologies ecosystem, practical and sustainable implementation of ACF must utilize AI augmentation to optimize performance. The fourth and final proposition of ACF is therefore articulated as:

**P4**. *Artificially intelligent cognitive fit solutions enable optimal performance by adaptively and dynamically aligning information facets, tasks and information representations.*

Such AI augmented ACF performance optimality can be expected to improve with time, as the algorithms learn not only to customize solutions to HI capabilities, but also to adapt to growth and changes to information facets and representations within the big data ecosystem.

*2.13. ACF hypotheses*

ACF hypotheses are based on ACF propositions and the theoretical arguments used to develop them. To demonstrate the influence of information facets on performance in the context of ACF, we operationalize equivocality in experimental settings, which is manifested using information scenarios. Such information scenarios engender equivocality by requiring participants to engage in complex tasks (e.g., trading stock to achieve profits subject to constraints), or manage ambiguous situations (e.g., resolving ethical dilemmas in organizational settings). The quality of HI decisions can be adversely affected by information with multiplicity of meanings. Therefore, for a given task, it is not only information representation, but also equivocality that can influence performance (Daft, 1987; Weick, 1979, 1995; Weick et al., 2005). Past research has shown that probabilistic and deterministic information representations have varying impacts on decision makers and performance (Webby & O'Connor, 1994; Chervany & Sauter, 1974; Samuel, Holowczak, & Pelaez, 2017). The development of an improved understanding of the role of equivocality in HI decision performance with deterministic and probabilistic information representations will lead to better decisions and predictions under varying levels of information equivocality (Kahneman, 2018).

*2.14. Equivocality – a dominant information facet*

Information facets such as equivocality and veracity can significantly alter HI's internal representation of information, thereby impacting performance (Grover et al., 2006). Equivocality of information is commonly understood as the degree to which an information scenario can reflect two or more meanings for HI (Wang, Sun, Shen, & Zhang, 2018). From an organizational perspective, equivocality has been defined as the "existence of multiple and conflicting interpretations about an organizational situation" (Daft & Lengel, 1986). Equivocal information possesses multiplicity of meanings, which can lead to varying conclusions or states of cognitive dissonance. The level of equivocality is the degree to which a variety of meanings can be associated with a given information artifact (Weick, 1979, 1995; Weick et al., 2005; Daft, 1987). Researchers have studied equivocality in choice performance (Dennis & Kinney, 1998), intuition (Julmi, 2019) and feedback (Brecher & Hantula, 2005). Equivocality has also been studied in the context of innovation processes (Frishammar et. al., 2010), strategic decision making (Neill & Rose, 2007), knowledge integration (Wang et al., 2018), and grammatical analysis (Hammerly, Staub, & Dillon, 2019). Higher levels of equivocality make information processing difficult, as they challenge the cognitive judgment and memory limits of HI (Miller, 1956). We posit that high levels of equivocality will have an adverse impact on performance.

### 2.15. Deterministic and probabilistic information representation

Probabilistic and deterministic information representations were contrasted at varying levels of equivocality, with the expectation that a lower level of equivocality would result in superior decision performance (Brecher & Hantula, 2005; Frishammar, 2010). In an experimental setting, Webby and O'Connor (1994) found that as task complexity increased, performance decreased, and participants exposed to probabilistic information took more time to make decisions. Similar comparisons were made in the "Minnesota Experiments" (Dickson et. al., 1977) by Chervany and Sauter (1974), between deterministic and probabilistic information representations with similar results. Further support for the deterioration in performance in project bidding processes was provided by Elmaghraby (1990), who reported higher project cost estimates from probabilistic information users (PERT) versus users of deterministic information (CPM). However, there is a scarcity of continued research on this interesting contrast of deterministic and probabilistic information management, which presented us with a unique and well aligned opportunity for this experimental study.

### 2.16. Equivocality and information representations

Our goal was to study the effects of high and low levels of equivocality upon participants exposed to deterministic and probabilistic conditions (Fig. 3). When there is conflicting information, participants can discover alternative values, but usually fail to associate likelihoods with these values. When a broader range of options are present, probabilistic information can provide multiple values (each associated with a likelihood of occurrence) related to a given asset and serve as a better representation of reality (Damodaran, 2007, 2009). Probabilistic information representations can support scenarios with high equivocality and were therefore expected to support high equivocality conditions better than deterministic representations.

We do not advocate the intrinsic superiority of one form of information representation over the other. Instead, we have attempted to understand the effects of varying information facets and information representations for a given task on performance. A significant performance difference for the same task, under different equivocality conditions, would validate the initial propositions of ACF and enhance our understanding of cognitive fit. Details of hypotheses development are discussed below, and the economic experiment is presented under the methods section.

### 2.17. ACF hypotheses

ACF focuses primarily on equivocality reduction and the interactions between information facets and information representations to improve performance. Leaning on dual information processing theory, we believe that low levels of equivocality engender the use of systematic processing. Based upon the 'least effort' principle of dual information processing, we suggest that deterministic information representation can lead to more effective performance under low equivocality. This is in contrast to higher levels of equivocality, when individuals facing higher cognitive challenges might be more prone to maximize the use of additional sensemaking mechanisms such as heuristics, schemas, and so on. As a result, probabilistic information representations, which can provide more options to reconcile equivocality induced dissonance, are more likely to improve performance under high equivocality. In summary, for the same task, we expect relatively better performance with deterministic representations under low equivocality in contrast to better performance with probabilistic models under high equivocality.

We have outlined the arguments as a basis to show that it is more critical to reduce equivocality for deterministic models in high equivocality conditions, than to reduce equivocality for probabilistic models in high equivocality conditions. Based on the above, we developed and tested five hypotheses: **Ha**, **Hb**, **Hc**, **Hd** and **He.** Hypothesis **Hb**, **Hc** and **Hd** can be viewed as emerging from hypotheses **Ha** and **He** – however, they need to be distinctly stated based on the logic that "*the whole is not always equal to the sum of the parts*": Though **Ha** posits that low equivocality will lead to better performance, we need to evaluate the individual contributions of deterministic and probabilistic information representations to the variance posited in **Ha**, as tested for in **Hb**, **Hc** and **Hd**. We emphasize that the first and last hypotheses (Ha and He) are most critical for a fair validation of ACF propositions P1 and P2.

Regardless of the type of information representation, a change in equivocality, from high to low levels, makes information processing easier for HI, and reduces the cognitive dissonance "burden", leading to better performance (Miller, 1956; Daft & Macintosh, 1981). Low equivocality also assists meaning construction (Cornelissen & Werner, 2014), and facilitates better understanding and interpretation of the information scenario (Lim & Benbasat, 2000; Zack, 2007). Therefore, we expect that:

**Ha.** *Subjects in low equivocality conditions will perform better than subjects in high equivocality conditions.*

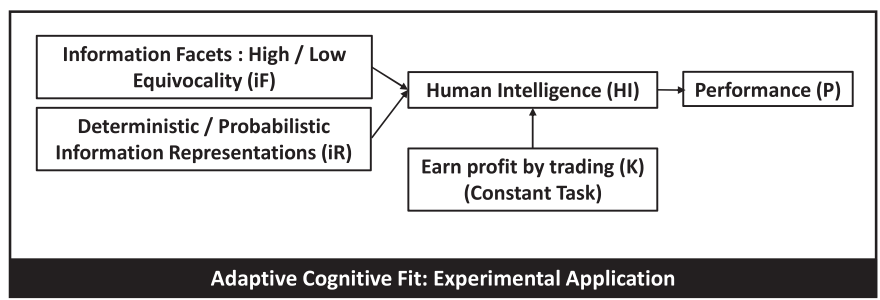

**Fig. 3.** Structure of ACF Experiment.

Extant research has demonstrated that when faced with choices, subjects choose to "minimize effort rather than maximize accuracy" (Benbasat & Todd, 1996). When there is conflicting information, participants can discover alternative values, but usually fail to associate likelihoods with these values due to the limits of their information processing capabilities. We expected subjects presented with deterministic information to experience higher cognitive dissonance under high equivocality due to conflicting perceptions emanating from the information scenario and the information processing limitations of HI. In contrast, subjects presented with probabilistic information under high equivocality would be able to utilize the probabilities to reconcile and decrease perceptual conflict to some extent, leading to a relatively lower cognitive burden (Daft & Macintosh, 1981). Therefore, subjects presented with a probabilistic information representation were expected to perform relatively better under high equivocality. This leads to the second hypothesis,

**Hb**. *Subjects presented with probabilistic information representation will perform better than subjects with deterministic information representation under high equivocality.*

Under low equivocality, subjects were expected to experience a lower cognitive dissonance than their counterparts under high equivocality, when exposed to probabilistic representation. In low equivocality conditions, cognitive dissonance is minimized temperately using probabilistic models. Hence, under low equivocality, subjects presented with probabilistic representation were expected to perform better than those under high equivocality with probabilistic representation. More formally, presented with a probabilistic information representation, subjects will perform better under low equivocality than under high equivocality:

**Hc**. *Presented with probabilistic information representations, subjects under low equivocality will perform relatively better than subjects under high equivocality.*

Under high equivocality, subjects were expected to experience a much higher cognitive challenge than their counterparts under low equivocality, when exposed to deterministic representation. Information processing load is the highest under high equivocality with deterministic representation. Under low equivocality, cognitive dissonance is minimized significantly with deterministic representation, and information processing becomes easier for HI. Hence under low equivocality, subjects presented with deterministic information representations were expected to perform significantly better than those under high equivocality. Therefore, more formally:

**Hd**. *Presented with deterministic information representations, subjects under low equivocality will perform relatively better than subjects under high equivocality.*

Previous research has shown that decision makers use heuristics, logic, and information in an adaptive manner to solve problems (Payne, 1982, 1988). ACF posits that changes in levels of information facets necessitate changes to information representations for the same task to improve performance. Therefore, we expected an interaction between the information representations and equivocality conditions in the determination of performance. We anticipated that while performance would improve under low equivocality, such improvement would depend upon potentially different information representation than under high equivocality. From an information processing standpoint, low equivocality conditions should increase the salience of relevant information, and lead to stronger performance improvement for deterministic representations than for probabilistic representations. Therefore, subjects presented with deterministic information under high equivocality would benefit significantly more from low equivocality than subjects presented with probabilistic information. Hence, we posit our critical hypothesis:

**He**. *The increase in performance of subjects in low equivocality versus subjects in high equivocality conditions, will be significantly greater for subjects presented with deterministic representations versus subjects presented with probabilistic representations.*

This implies that there will be *an interaction between equivocality and information representations.* The fifth and most important ACF hypothesis (He) reflects the critical need for the "adaptive" nature of cognitive fit presented in the ACF framework, wherein optimal decisions are achieved not just by aligning information representations to tasks, but by preceding alignments with adaptivity to the levels of information facets.

### 2.18. Methodology

We employed an economic experiment to test the hypotheses that are aligned with ACF propositions 1 and 2, in accordance with the call for increased use of this approach in IS research (Gupta, Kannan, & Sanyal, 2018). We validated ACF propositions 3 and 4 with machine learning simulations. Our economic experiment is well aligned with research methodologies in information systems and decision support systems, as well as behavioral economics and cognition (Arnott & Gao 2019; Smith 1962, 1976). The role of information representation in decision making has been studied using laboratory experiments (Chervany & Sauter, 1974) and experimental simulations (Webby & O'Connor, 1994). In our experiment, we used an interactive equity market simulation in a laboratory setting to study the influence of equivocality and information representation upon performance. Our goal was to explore the influence of equivocality upon trading performance.

### 2.19. Experimental setup

We used "TraderEx", a stock trade simulation and market microstructure management training platform (Fig. 4), for the experiment (Schwartz, Sipress, & Weber, 2010; Marriott, Tan, & Marriott, 2015). Subjects were allowed to trade one stock in an equity market simulation under varying levels of equivocality induced by information provided through a custom designed website (Appendix A, Fig. 2). The trading simulation deployed a single equity electronic market to avoid subjects experiencing confounding effects from observing the price movements of multiple equities. We used a between-subjects 2 conditions by 2 conditions research design, consisting of high and low equivocality information artifacts in the form of relevant news articles, and deterministic and probabilistic information representations presented to the subjects using the website (Appendix A, Figs. 2,3,4 and 5). Information content was not varied significantly, except for the format of the content as was required to stage the experimental conditions. We conducted a manipulation check (described below) for the experimental artifacts and then ran the main experiment. Subjects were required to trade for profit, that is to buy and sell stock, subject to the experimental conditions, and earn profits. Subjects were given a starting price for the stock, and dividend information about the stock. They were trained to use the deterministic and probabilistic representations to gain an approximate but fairly accurate end of day trading price. Performance was measured by the end-of-day trading profit earned by each subject. The hypotheses were tested using ANOVA to study the effects of equivocality for the various information representations on performance.

### 2.20. Domain logic and experimental validity

Stock prices have traditionally been believed to demonstrate random behavior, especially in the short term, but subsequent research has also identified the existence of short-term patterns under a variety of market microstructure and investor behavior conditions (Fama, 1965; Lo, 2019). Extant research has confirmed the presence of intraday stock price patterns and trends, under varying market conditions such as bull and bear market phases, and investor sentiment behavior (Kottaridi,

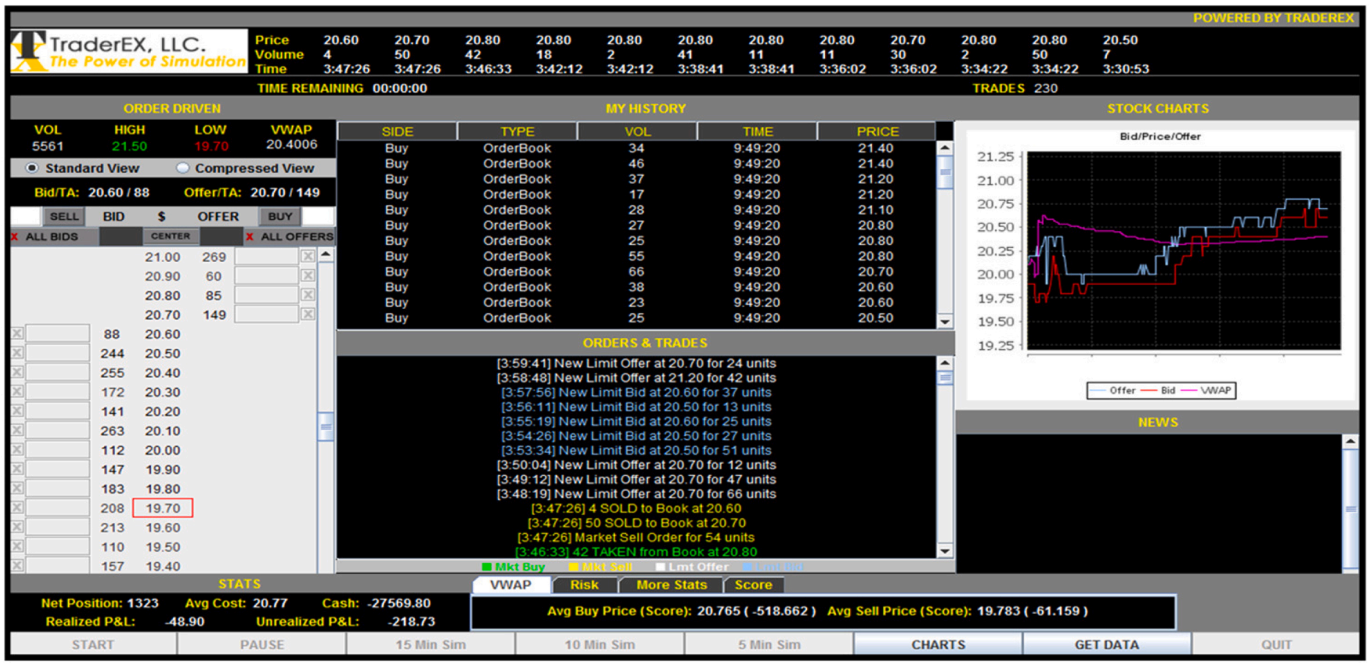

**Fig. 4.** TraderEx trading platform.

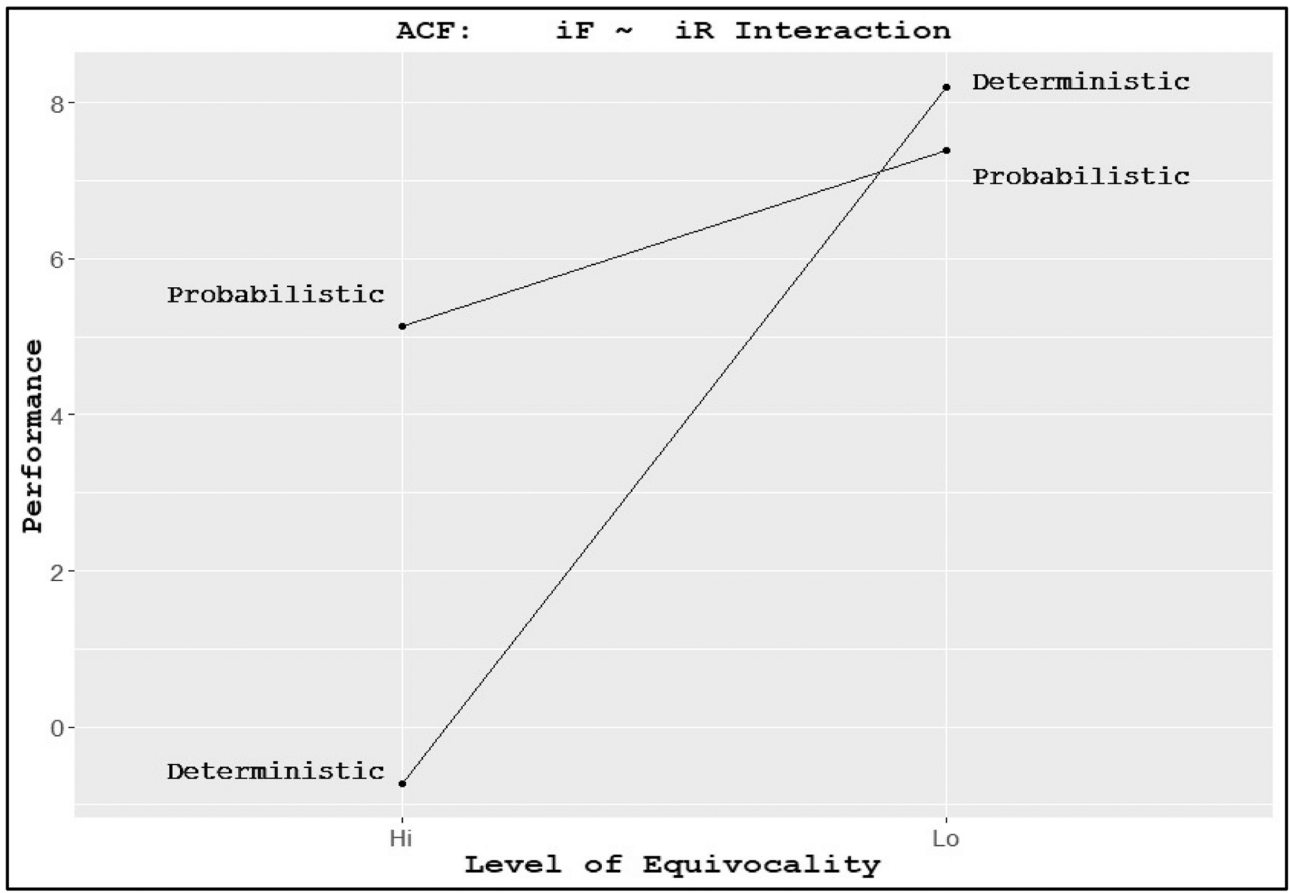

**Fig. 5.** Equivocality Levels and Model Types Interaction Plot.

Skarmeas, & Pappas, 2020; Alexakis, 2003; Renault, 2017). In the present study, we use an information driven stock price scenario to simulate the single stock experimental equity market, which is in line with market scenarios under new information such as news about the development of a new product or unusual increase in sales or similar positive or negative news (Feuerriegel, Ratku, & Neumann, 2016; Allen, McAleer, & Singh, 2019). The between-subjects experiment provided textual information with high and low equivocality to the subjects using a custom website (Appendix A, Fig. 2), and the news corresponded to a preset increase in the stock price on TraderEx, mixed with volatility and noise, as would be expected in a stock market.

Subjects were required to take the information (high or low equivocality) and apply a deterministic or probabilistic model provided to them to make an informed trading decision with the goal of making a profit by buying and selling the stock. Subjects had the option to short-sell the stock as well. This ensured that the subjects had reasonable means to make a profit through informed trading, and most importantly, the experiment controlled all variables and only the four experimental conditions were varied: *high-equivocality by deterministic and probabilistic models, and low-equivocality by deterministic and probabilistic models.*

### 2.21. Manipulation check

We used manipulation checks to verify the effectiveness of the experimental conditions and ensure internal validity of the experiment. We based the manipulation check on Hantula's (1999) manipulation check for equivocality conditions (Appendix A, #4). Dennis and Kinney (1998) also used a similar manipulation check which they derived from Daft and Mcintosh (1981) and Daft and Lengel (1986). The manipulation checks confirmed the effectiveness of the experimental conditions for equivocality reduction by measuring how the subjects differentiated between the levels provided in the experiment. We induced high and low equivocality levels using textual information displayed on the website, and deterministic and probabilistic information representations using valuation models (Appendix A, #5).

Probabilistic and deterministic models are determined by definition, while equivocality is perceived and so the manipulation check was focused on validating the difference in high and low equivocality, under each information representation (Dennis & Kinney, 1998). The manipulation check included the following items: a) texts with high and low equivocality, with a probabilistic model and b) texts with high and low equivocality, with a deterministic model. The goal was to ensure that the experimental artifacts served the purpose of varying equivocality under both deterministic and probabilistic conditions. A 7-point Likert scale survey was conducted to study the effects of the experimental conditions (Appendix A, #4). The results of the manipulation check are posted in Tables 2 and 3. Manipulation check A (Table 2) based means for low versus high equivocality for the probabilistic condition were 1.62 and 5.81 and manipulation check B (Table 3) based means for low versus high equivocality for the deterministic condition were 1.53 and 6.05 respectively. The differences in means between high and low equivocality conditions for both information representations were found to be highly significant ($p < 0.001$). The manipulation check confirmed the effectiveness of the experimental artifacts and affirmed the internal validity of the experiment.

**Table 1**
List of main abbreviations used to present this study.

| Abbreviation | Phrase / Term | Abbreviation | Phrase / Term |
|---|---|---|---|
| ACF | Adaptive Cognitive Fit | iF | Information Facet |
| AI | Artificial Intelligence | iR | Information Representation |
| CFT | Cognitive Fit Theory | K | Task |
| ML | Machine Learning | P | Performance |
| HI | Human Intelligence | GUI | Graphic User Interface |

**Table 2**
Manipulation Check A.

| Equivocality ~ Probabilistic | |
|---|---|
| t = -14.69, df = 31.10<br>Means: Low-Eqv and Hi-Eqv<br>1.62 and 5.81<br>Welch Two Sample t-test | p < 0.001 *** |

Significance Codes: 0 '***' 0.001 '**' 0.01 '*' 0.05 '.' 0.1

**Table 3**
Manipulation Check B.

| Equivocality ~ Deterministic | |
|---|---|
| t = -22.23, df = 35.48<br>Means: Low-Eqv and Hi-Eqv<br>1.53 and 6.05<br>Welch Two Sample t-test | p < 0.001 *** |

Significance Codes: 0 '***' 0.001 '**' 0.01 '*' 0.05 '.' 0.1

### 2.22. Experimental procedure

The experiment used eighty-five student subjects from the CIS student subjects-pool at a large US university. They were recruited through the school's research subject enrollment system. Subjects were undergraduate students with no prior exposure to electronic stock trading. We used the TraderEx trading simulation platform which is used for teaching and training on market microstructure topics by graduate and undergraduate students and professionals. This trading simulation platform is dissimilar to generic market platforms, ensuring a level playing field for the participants. Subjects were given a demonstration of the trading software, followed by two practice rounds to familiarize themselves with the use of the TraderEx platform. The main round was a 15-minute simulation of one trading day – the simulation clock started at 9:30 AM and moved to 4:00 PM in 15 minutes. After the practice rounds, subjects were introduced to equivocality and model conditions (Table 4) in different sessions and tasked to trade with the objective of maximizing realized profits by buying low and selling relatively higher. To perform this task, they had to deal with high or low levels of equivocality and deterministic or probabilistic conditions, to determine the direction the stock price would move in. In line with the principles used in economic experiments (Smith, 1976), a performance-based cash payout was announced for the top performers to induce appropriate motivation to trade for economic benefits. All students received course credit for their undergraduate computer information systems course for participating in the experiment.

Following the verifications from the manipulation checks, the experimental design was finalized as a 2×2 between-subjects experiment set up for a two-way ANOVA (Appendix A). The experiment was based on a single decision maker framework. Using Table 4, the experiment simulated information equivocality conditions. Equivocality levels were varied as high and low equivocality conditions under deterministic and probabilistic model conditions to explore changes in performance. Changes in equivocality levels and model types were expected to affect performance (P). The operational conditions of the experiment are schematically represented above (Table 4). C1 represents high equivocality under deterministic representation, C2 represents high equivocality under probabilistic representation, C3 represents low equivocality under probabilistic representation, and C4 represents low equivocality under deterministic representation. Subjects were introduced to these conditions on a between-subjects basis after the demonstration and practice rounds. They were asked to trade to maximize their end-of-day trading profits based on the information provided. At the end of the experiment, performance was recorded for each subject, subjects were debriefed and top performers in each condition received performance-based cash rewards. We controlled for three conditions, which included prior use of the software platform, prior equity trading experience and gender. None of the participants indicated prior usage of the trading simulation platform or prior trading experience, and so there was no need to include those variables in the analysis. The study included 40 female subjects and 45 male subjects. We controlled for gender and did not find any indication that gender influenced performance.

**Table 4**
Experimental conditions mapped to the ACF Framework.

| Information Representation (iRT) | Information Facet ( iF) | |
|---|---|---|
| | High Equivocality | Low Equivocality |
| Probabilistic Model | C2: Low Performance | C3: Higher Performance |
| Deterministic Model | C1: Lowest Performance | C4: Highest Performance |

*2.23. Analysis and results*

All hypotheses (**Ha, Hb, Hd** and **He**), except **Hc**, were supported by the results of the data analysis. The most insightful result is that the interaction hypothesized in **He** between equivocality and information representation is supported by the experiment (Fig. 5). The descriptive summary is presented in Table 5 and the main results are presented in Table 6 to 7. We used descriptive statistical analysis and ANOVA to study the data.

We analyzed the data anticipating support for a critical ACF hypothesis Ha, which posits that sufficient variance in levels of equivocality would influence performance significantly. Hypothesis **Ha** was supported and subjects under low equivocality conditions performed better than subjects under high equivocality conditions ($P < 0.01$; Table 6). We believe that this intrinsically represents an important finding, as it demonstrates the power of information facets to influence performance. Table 6 represents aggregate level analysis. This validates the experiments at the level of information facets, but there was a need to analyze the components in greater detail to separately study the effects of information representations under high equivocality (Hb), the effects of probabilistic representation under high and low equivocality (Hc), and the effects of deterministic representation under high and low equivocality (Hd). This was possible because the experiment was designed using a between-subjects arrangement, which ensured the independence of each experimental condition. Hypotheses Hb, Hc and Hd are analyzed at a more granular level in the following section (Table 7A, 7B, 7C).

**Hypothesis Hb.** was supported and subjects presented with probabilistic information representation performed better than subjects presented with deterministic information representation under high equivocality ($P < 0.05$; Table 7A). The average performance under the deterministic condition was -0.72, and under the probabilistic condition was 5.13, under high equivocality. Given that we hypothesized this effect only under high equivocality, we ran the ANOVA as a one-way analysis for information representation means, positing that there is a significant difference in performance means between deterministic and probabilistic conditions.

**Hypothesis Hc.** was not supported at an alpha of 0.05. Though subjects presented with probabilistic information representation, performed better under low equivocality than subjects under high equivocality, and the results were moderately supported with a P value of 0.07 ($P < 0.1$, Table 7B). We were not completely surprised as it was plausible to envision the subjects using both probabilistic models under high equivocality achieving a fair degree of success in completing the trading task, and hence not showing significantly large improvement under low equivocality.

**Hypothesis Hd.** was supported and subjects presented with deterministic information representation under low equivocality performed significantly better than subjects presented with deterministic information representation under high equivocality ($P < 0.01$; Table 7C). Performance means were -0.72 and 8.19 for high and low equivocality, respectively ($P < 0.01$; Table 7C). Since we hypothesized this effect only for deterministic representations, we ran the ANOVA as a one-way analysis for equivocality performance means, positing a significant difference in performance means based on equivocality levels (Table 7C).

The most critical hypothesis **He** was unequivocally supported by the analysis ($P<0.05$, Table 6, Fig. 5). This demonstrates the presence of an interaction effect between equivocality and information representations. Hypothesis **He** was supported and the average positive difference in performance between-subjects under low equivocality and subjects under high equivocality, was significantly greater for deterministic information representation than probabilistic information representation, confirming the face value of the interaction effects ($P < 0.05$, Table 6, Fig. 5). Table 6 validates the most insightful ACF hypothesis (**He**): the interaction between information facet (equivocality) and information representation (deterministic and probabilistic). The analysis presented in Table 6 also indicates weak support for information representations at an aggregate level, at P value = 0.06 (Table 6). However, we find it noteworthy that under high equivocality, the difference between deterministic and probabilistic representations is significant with a P-value of 0.03 ($P < 0.1$; Table 7A). Therefore, this aspect merits further investigation with alternate tasks and information representations

We employed three control variables: prior use of the TraderEx software platform, prior equity trading experience and gender. We only

**Table 6**
ANOVA for performance means of high and low equivocality.

| Equivocality and Information Representations ANOVA | | | | | |
|---|---|---|---|---|---|
| | Df | Sum sq | Mean Sq | F value | Pr (>F) |
| Equivocality | 1 | 629 | 629.3 | 15.02 | 0 *** |
| Information Representations | 1 | 148 | 148.5 | 3.54 | 0.06. |
| Equivocality x Information Representations | 1 | 237 | 237.1 | 5.66 | 0.02 * |
| Gender | 1 | 1 | 0.6 | 0.02 | 0.9 |
| Residuals | 80 | 3352 | 41.9 | | |

Significance codes: 0 '***' 0.001 '**' 0.01 '*' 0.05 '.' 0.1

**Table 5**
Descriptive Analysis: Summary of Performance Means and Standard Deviations.

| iF | Mean (P) | Std. Deviation | Under Low Equivocality | | |
|---|---|---|---|---|---|
| High-Equivocality | 2.34 | 9.16 | iR | Mean (P) | Std. Deviation |
| Low-Equivocality | 7.78 | 1.79 | Deterministic | 8.19 | 2.39 |
| iR | Mean (P) | Std. Deviation | Probabilistic | 7.39 | 0.81 |
| Deterministic | 3.63 | 9.31 | | | |
| Probabilistic | 6.21 | 4.21 | Under High Equivocality | | |
| Gender | Mean (P) | Std. Deviation | iR | Mean (P) | Std. Deviation |
| Female | 5.17 | 5.85 | Deterministic | -0.72 | 11.27 |
| Male | 4.77 | 8.3 | Probabilistic | 5.13 | 5.61 |

**Table 7A**
ANOVA for performance means under high equivocality.

| High Equivocality: Deterministic and Probabilistic iR | | | | | |
| --- | --- | --- | --- | --- | --- |
| | Df | Sum sq | Mean Sq | F value | Pr(>F) |
| iR: Det and Prob | 1 | 377 | 376.8 | 4.9 | 0.03 * |
| Residuals | 42 | 3233 | 77 | | |

Significance Codes: 0 '***' 0.001 '**' 0.01 '*' 0.05 '.' 0.1.

**Table 7B**
ANOVA for performance means under probabilistic condition.

| Probabilistic iR: High and Low Equivocality | | | | | |
| --- | --- | --- | --- | --- | --- |
| | Df | Sum sq | Mean Sq | F value | Pr(>F) |
| iF: High and Low | 1 | 56 | 56.03 | 3.34 | 0.07. |
| Residuals | 42 | 704.7 | 16.78 | | |

Significance Codes: 0 '***' 0.001 '**' 0.01 '*' 0.05 '.' 0.1

analyzed gender because no subject indicated prior TraderEx platform usage or prior trading experience. Age was not a concern due to the largely homogenous (by age) subject pool. Performance means were mostly similar between male subjects and female subjects (Table 5). Our analysis of the performance of 40 female subjects and 45 male subjects did not indicate any gender influence on performance (Table 6). Although female subjects had better performance means under both high (F= 2.52, M= 2.19) and low equivocality (F=7.83, M=7.74), and overall performance means (F= 5.17, M= 4.77), Table 6 indicates that gender-based performance differences were not significant. The two most critical ACF Hypotheses, Ha and He, were firmly supported, validating the importance of factoring information facets into cognitive fit for an improved understanding of human performance.

We took several precautions to minimize threats to internal validity. All groups were briefed by the same researcher to avoid any potential differences in researcher characteristics or communication styles to influence outcomes. The trading simulation deployed a single equity electronic market to avoid any potential confounding effects that may have arisen from observing the price movements of multiple equities. We provided balanced information artifacts to all groups to avoid inferences about equivocality being drawn from the volume of text or appearance of the information artifact. We had a rule to exclude subjects with prior trading experience to avoid confounds based on differences in expertise. We conducted manipulation checks (as reported earlier) to confirm high and low equivocality. Subjects in all conditions were provided with the same training on the use of valuation models.

We also took various measures to minimize threats to external validity. First, we deployed TraderEx, which provides a realistic stock trading environment and is used to train students and traders. We deployed an end-of-day trading profit measure to compare performance between experimental conditions which is reflective of what happens in the practice of day trading in stock markets. A website was developed for the purpose of displaying information in the form of news announcements to strengthen the external validity of the information artifacts. We also provided performance based cash payouts to all subjects, which is the established approach in research using economic experiments, and is reflective of the performance based rewards traders would experience in real-world trading.

**Table 7C**
ANOVA for performance means under deterministic condition.

| Deterministic iR: High and Low Equivocality | | | | | |
| --- | --- | --- | --- | --- | --- |
| | Df | Sum sq | Mean Sq | F value | Pr(>F) |
| iF: High and Low | 1 | 814.5 | 814.5 | 11.99 | 0.00 ** |
| Residuals | 39 | 2650.1 | 68 | | |

Significance Codes: 0 '***' 0.001 '**' 0.01 '*' 0.05 '.' 0.1.

### 2.24. ACF: AI application

Simulation methods and adaptive programming have been adopted in recent IS research to develop and validate theoretical propositions (Pentland, Liu, Kremser, & Hærem, 2020). We designed a simulation that uses artificially intelligent learning methods to validate ACF propositions 3 and 4. Unsupervised machine learning, a popular AI method, was deployed to accurately select the best-fit information representations. The objective was to train a ML model which could classify performance by the four experimental conditions: high and low equivocality, and deterministic and probabilistic conditions, with sufficient accuracy (our target: 95% accuracy). Once a ML model accomplishes this objective, an information systems solution can provide a machine generated selection of information representations to HI (Fig. 6). Not only does this illustrate the potential for AI enabled ACF design solutions, but it also offers insights to addressing challenges posed by big data with broader information systems designs.

We used a simulation to generate data aligned with the behavior of human subjects as seen in our human subjects-based experiment (Fig. 7). An unsupervised machine learning method was used to cluster the simulated performance values and analyze the levels of accuracy achieved. Once sufficient accuracy was realized, the unsupervised ML provided output which indicated the best fit information representation for optimal performance with over 98% accuracy. From this point, design solutions could either automate AI-selected representations or provide AI recommendations for HI (dotted lines, Fig. 6), thus illustrating the value of the final propositions of ACF (Fig. 6, Tables 8, 9). We used R, one of the leading analytics-oriented statistical computing languages, to simulate data using a pseudo-random number generator and a normal-distribution function to represent the performance associated with the four experimental conditions (R Core Team, 2018; RStudio, 2020). We simulated 4000 data points, modeled based on human behavior observed in the experiment, in the simulation, including 1000 data points for each condition, in alignment with the human subjects' experiment (Figs. 5 and 7). Subsequently, we used R packages for the unsupervised machine learning application and accuracy metrics. We validate this discussion with a simple design schema to demonstrate the feasibility of using AI for managing information representations (Fig. 6).

### 2.25. ML simulation

Recent IS research has leveraged the analytics-oriented language R, and associated tools and packages for machine learning driven healthcare information systems research (Son, Flatley Brennan, & Zhou, 2020). We simulated data to represent the performance associated with the four information-facet and model type conditions using the R software, and subsequently used R packages for applying machine learning. The behavior of the simulated data aligns with the behavior of data in our experiment with human subjects, as shown in Fig. 7. Four thousand data points were generated, with one thousand data points for each condition, using a pseudo-random number generator in R, and applying a normal-distribution function as the baseline for the simulation process. We used the Mclust package in R to apply unsupervised machine learning to the simulated data without labels that might indicate the level of equivocality or the information representation (Scrucca, Fop, Murphy, & Raftery, 2016). The clustering algorithm generated a Gaussian finite mixture model fitted by an EM (Expectation-maximization) algorithm with four components of 1000, 993, 1007, and 1000 values in each of the four clusters. Multiple models were generated using standard model-based agglomerative hierarchical clustering processes, using both original variables and scaled SVD (singular value decomposition) transformations to optimize model selection. The best model was determined based on the best BIC (Bayesian information criterion) score as shown in Fig. 8.

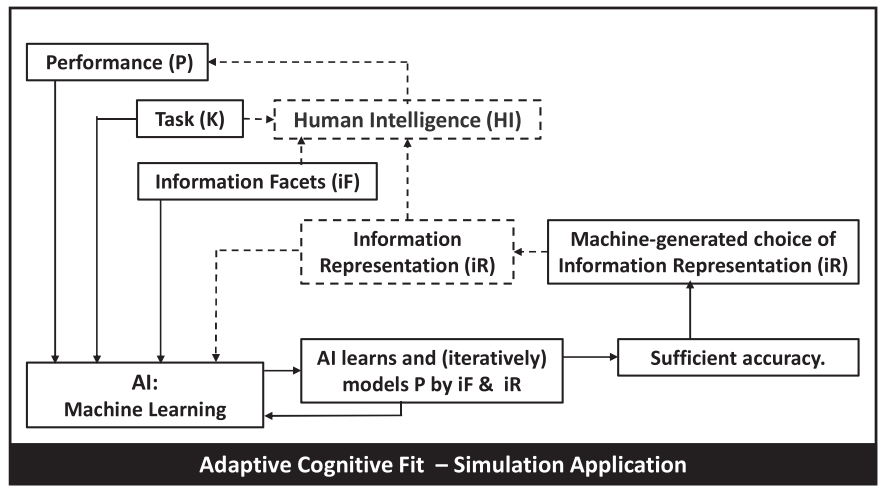

**Fig. 6.** Applied ACF design schema for simulation with machine learning.

*2.26. ML analysis*

We generated a confusion matrix by comparing the artificially generated clusters from the simulation with the naturally occurring labels for the four 'information-facet x information-representation' conditions, to study the accuracy of the model as shown in Tables 8 and 9. Overall Accuracy of the model is given by True Classification / Total Classification, that is (1000 + 968 + 975 + 1000)/4000, which equals an overall accuracy of 98.575%. We also developed precision and accuracy metrics to validate classification reliability measures for the overall classification, and by class, as displayed in Table 9 A and B, respectively.

*2.27. ACF design schema*

The results presented in the confusion matrix demonstrate a high degree of accuracy provided by the machine learning generated model. This parsimoniously illustrates and validates the feasibility of using artificially intelligent learning methods to map human performance to models of information facets and information representations (Fig. 6). The straight lines in Fig. 6 represent the components illustrated by the ML simulation above. The dotted lines indicate options for AI-recommendation or AI-implementation, and the figure in its entirety represents the full design schema for artificially intelligent ACF. The complete ACF schema could be implemented as an information systems

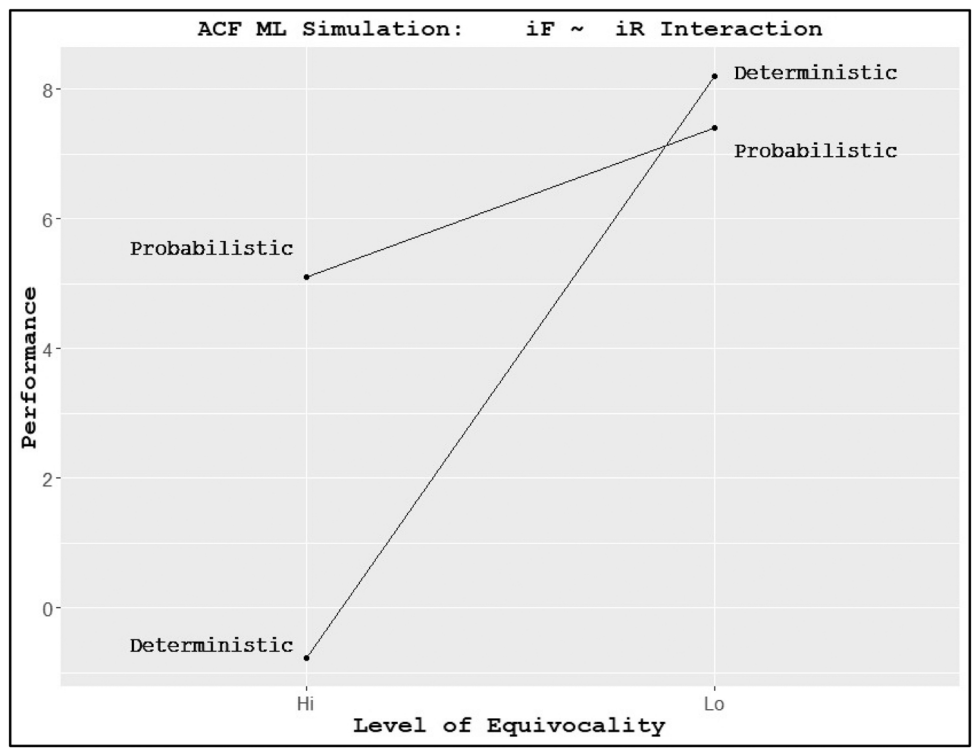

**Fig. 7.** Simulated information facets and representations interaction.

**Table 8**
Confusion Matrix for Simulated Information-facets: Representations.

| Class | 1 | 2 | 3 | 4 |
|---|---|---|---|---|
| **HiEqv-Det** | **1000** | 0 | 0 | |
| **HiEqv-Prob** | 0 | **1000** | 0 | 0 |
| **LoEqv-Det** | 0 | 0 | **968** | *32* |
| **LoEqv-Prob** | 0 | 0 | *25* | **975** |

**Table 9**
A, B: F1, Accuracy, Precision and Recall Metrics.

| Table: 9 A | **Overall** | | | |
|---|---|---|---|---|
| **Accuracy** | 98.58% | | | |
| **Precision** | 98.58% | | | |
| **Recall** | 98.58% | | | |
| **F1** | 98.57% | | | |
| **F1\* (Weighted)** | 98.57% | | | |
| Table: 9 B | HiEqv-Det | HiEqv-Prob | LoEqv-Det | LoEqv-Prob |
| Cluster: | 1 | 2 | 3 | 4 |
| Precision | 100.00% | 100.00% | 97.48% | 96.82% |
| Recall | 100.00% | 100.00% | 96.80% | 97.50% |
| F1 | 100.00% | 100.00% | 97.14% | 97.16% |

(* Weighted scores remain the same as there are 1000 data points for each class)

application solution which combines 1) observations of information facets, human performance and all associated variables, 2) ML algorithms to model information facets, information representations and tasks, leading to adaptive selections of information representations for optimal performance and 3) the optimization loop, which feeds the machine generated information representation forward to HI, and feeds back corresponding performance to keep improving the ML process and achieve optimal adaptive cognitive fit. Optimization can be used at multiple points, contingent on design requirements, in the ACF design schema to support the ML process, and to accommodate updates to goals, resources and constraints when the ML process has achieved sufficient accuracy.

Our objective was to train a ML model to accurately classify performance into the four experimental conditions, created by high and low equivocality with deterministic and probabilistic information representations. As this was successfully implemented (Table 9 A, B; Figs. 6 & 7), we suggest that in the future, IS designers should consider using ML for generating artificially intelligent selections of information representations to HI. By conducting this simulation using machine learning, we have demonstrated the potential for aggregate level AI augmented performance improvement (Cavalcante et al., 2019). The same conceptual model can be adapted for AI augmentation at the individual level, with sufficient training data and iterations, highlighting the potential for personalized AI driven performance improvement. Not only does this illustrate the potential for AI enabled ACF design solutions, but it also offers insights to address the challenges posed by big data with adaptive information systems designs as a whole.

## 3. Discussion

Big data and AI are driving radical technological change at many levels - from self-driving cars to facial recognition, machine translation, robotics and conversational AI, to name a few. Illustratively, factories "*are using context-aware "cobots" (collaborative robots) with "exoskeletons": wearable robotic devices attached to industrial workers that adapt quickly to the workers and their location, and function in an integrated manner with the workers to perform jobs with superhuman endurance and strength*" (Rai et al., 2019). We believe that it is in the context of such emerging AI ecosystems that ACF provides profound thought leadership and practitioner value creation potential through artificially intelligent cognitive fit solutions that augment human performance.

ACF provides a framework for "*seeking optimal outcomes under uncertainty*" in scenarios which require information processing under the cognitive challenges of dominant information facets such as equivocality (Baird & Maruping, 2021). The outcomes provisioned by ACF are distinct from the outcomes of traditional cognitive fit and other non-adaptive decision-making frameworks that do not include the influence of information facets. ACF based outcomes demonstrate sensitivity to changes in information facets even when there is no change in the form of the information artifact representing a flow of information. For example, commodity prices could be represented by graphs, with or without equivocal information. Furthermore, the ACF framework provides a basis to study interaction effects between latent information facets and associated information representations, such as graphs or tables or text or hybrid, complex and dynamic big-data driven visualizations and representations. The identification of such interaction effects is a unique outcome of the application of the ACF framework which can be very valuable for several applications, including human-machine interface design. Finally, ACF improves performance outcomes, and with sufficient data and training iterations, ACF can lead to best possible or optimal outcomes.

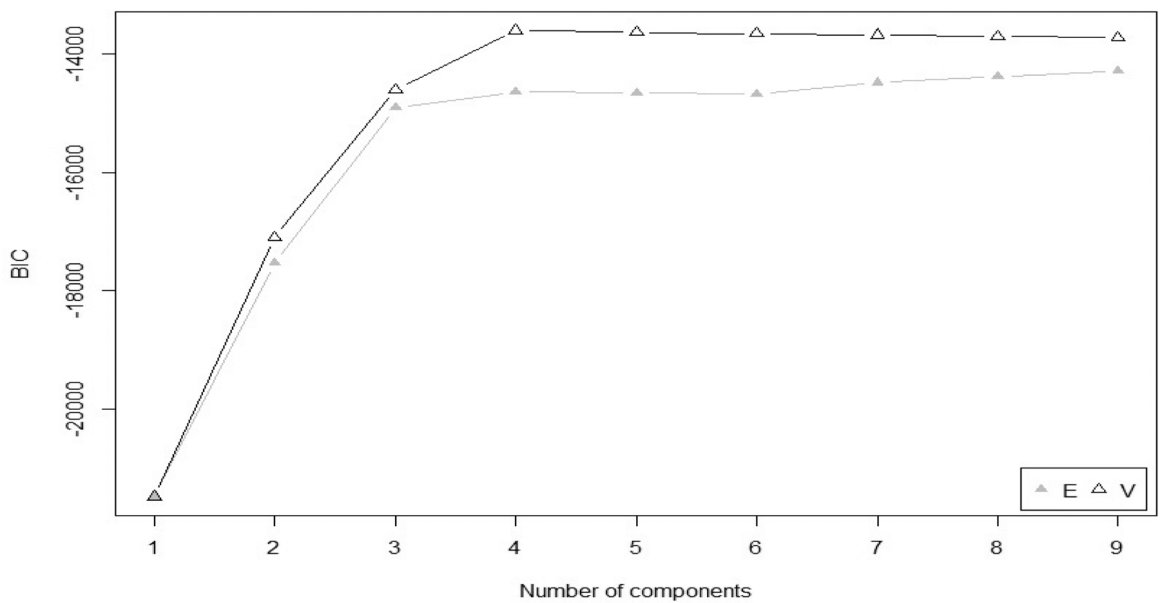

**Fig. 8.** Algorithmically generated clusters by BIC score.

A key contribution of ACF is the recognition of the variance of information facets, and the influence thereof on human performance. Furthermore, ACF is task agnostic, so that the influence of information facets on performance can be studied, while varying or keeping task constant. The effects of information representations can be both subjective and temporal under the influence of varying information facets such as equivocality. ACF posits the need for and viability of progressive adaptation of information representations to counter the challenges posed by variations (i.e., extremity, intensity) of information facets through AI augmentation to improve performance. The ACF framework facilitates AI augmented information representations driven optimal cognitive fit. ACF is distinct from CFT as it accommodates the information properties magnified by big data, analytics and AI technologies, such as equivocality and uncertainty. The application potential of ACF is also significantly different from CFT, as ACF integrates the use of artificial intelligence methods and technologies.

*3.1. Theoretical contributions and implications*

ACF has significant theoretical implications and presents several opportunities for future theoretical development. The critical concepts of the adaptive nature of cognitive fit, the influence of information facets and the self-learning capabilities of artificial intelligences converge under ACF to serve as a new theoretical lens for cognitive fit research contextualized to the increasing complexities of modern-day information environments. We as the human race have crossed certain thresholds in the development of information artifacts' volumes and dynamism which compel us to seek AI augmentation to overcome HI's information processing limitations made obvious by the very information ecosystems we have created. ACF, in addition to the direct application potential it possesses, can be used for future theoretical extensions to multiple streams of research such as cognitive fit, AI augmentation for human performance improvement, and the design of adaptive systems.

In this section, we summarize and highlight our reasons for believing that ACF can be extended to several significant streams of research: **First,** ACF specifically addresses the challenges of the new and evolving information environment. Therefore, ACF can help IS researchers to study ways to overcome HI's information processing limitations in coping with the escalating range and complexities of information artifacts. For instance, research may be directed at developing IS frameworks and artificially intelligent systems designs for autonomously triggering AI augmentation in the presence of significant changes to or dominance of information facets. Such inquiry can guide the design of information systems to manage HI-IF interactions and performance optimizations in specific domains, such as medical imaging analysis. **Second,** ACF is task agnostic (performance can be influenced with or without change in tasks). It focuses on the changes in human performance as a result of changes in information facets. ACF implies that the fit between tasks and information representations needs to be dynamic and adapted to HI through AI augmented user experiences that ease cognitive loads. Such ACF driven AI augmentation is capable of learning from both individual HI and group level information processing and performance Therefore, ACF can be extended to research the effects of adaptive systems on individual and on group tasks. ACF thus provisions opportunities to develop adaptive decision support systems for individual use and in organizational settings. **Third,** ACF can catalyze future cognitive fit research, supporting the integration of concepts from areas such as cognitive mapping, as it theoretically demonstrates the vanguard importance of information facets, emphasizes the need to evaluate multiple information dimensions, and highlights the necessity of AI augmentation from a design science perspective. For instance, by enhancing the scope for cognitive mapping of information facets, ACF opens the door for machine learning based predictive analysis.

Researchers can deploy ACF and test AI augmented information representations to improve learning and forecasting accuracy.

**Finally,** ACF provides a new lens to reframe past cognitive fit studies in the context of the big data ecosystem and examine risks through the study of lurking influences of information facets such as equivocality and uncertainty. This study adds value to the body of cognitive fit research in information systems and decision support systems, by calling HI, AI, technology and information artifacts simultaneously into focus (Orlikowski & Iacono, 2001; Chen, Chiang & Storey, 2012; McKinney & Yoos, 2010). The ACF framework can help researchers study the influence of other information facets such as uncertainty, veracity and overload, and their interactions with information representations on cognitive fit. For instance, drawing on extensive research in the social sciences about decision making under uncertainty, researchers can analyze differences in the effects of uncertainty on HI under varying conditions of AI augmented information representations.

Artificial Intelligences, using machine learning in particular, are being used for a wide range of adaptive solutions to support human intelligence with information source selection, digital agents, intelligent robotics, CaVs, textual analytics and natural language processing and generation (Lebib, Mellah, & Drias, 2017; Baird & Maruping, 2021; Samuel, Ali, Rahman, Esawi, & Samuel, 2020; Khayyam, Javadi, Jalili, & Jazar, 2020; Garvey, Samuel, & Pelaez, 2021; Samuel, Palle, & Soares, 2022). We anticipate ACF will stimulate additional inquiry into such areas, which involve HI interactions with complex information scenarios, and support research streams for AI augmented cognition.

*3.2. Implications for practice*

We believe that ACF is a significant departure from the long and valuable history of past cognitive fit research which has focused on manual mapping of specific information representations to specific tasks (i.e., Vessey, 1991 to Walden et al., 2018). ACF resolves the practical implementation challenges of manual approaches to determine cognitive fit in the context of big data, resulting complexities and an ever-increasing array of information representations. ACF does not posit specific information representations solutions for tasks. Instead, ACF demonstrates that cognitive fit must use AI methods such as machine learning to dynamically adapt information representations to information facets to produce superior and optimal performance. This will enable artificially intelligent information systems to generate affordances by creating adaptive solutions at the individual and collective (group, teams or organizational) levels. Specific ACF implications are presented below:

**First,** from a practitioner perspective, ACF accommodates the variances in information facets, information representations, tasks and human intelligence to maximize performance. This could be applied to a broad range of scenarios, including stock trading, medical imaging and gaming. ACF logic opens avenues for a wide range of applications such as the dynamic and automated selection of appropriate data representations. **Second,** ACF based architecture and designs can be used to increase user satisfaction through fit-personalization. AI augmented solutions can adapt to and map individual and collective HI capabilities. For example, gamification solutions can use ACF to match HI to GUIs (graphic user interfaces) to improve outcomes (Liu, Santhanam, & Webster, 2017).

**Third,** ACF has the power to mitigate the dissonance of individual perceptions due to changing levels of information facets in an organizational setting. Real-world information is complex, and subject to multiple and often conflicting interpretations by HI (Oh, Agrawal, & Rao, 2013). ACF can facilitate the identification of relevant information facets to enable AI design solutions for optimal performance. **Fourth,** ACF provides critical insights into human behavior under equivocality conditions, which are extensible to other information facets. With additional domain specific research, practitioners can apply the principles of ACF to the design of decision support dynamic GUI, by adaptively aligning GUI design to dominant information facets. For example, eCommerce systems can be designed to adaptively display GUI which maximizes a suitable performance measure such as amounts traded or satisfaction.

**Fifth,** ACF can improve the likelihood of systems success, which is traditionally driven by information quality (Petter, DeLone, & McLean, 2008). ACF demonstrates the potential for automated improvement of relative performance under challenging information conditions such as high equivocality. Practitioners can develop artificially intelligent solutions for improving relative performance in spite of information quality issues of understandability and veracity. The ACF experimental study demonstrates the value of addressing information facets levels, such as equivocality, over the use of fixed information representations. Practitioners and organizations can apply ACF propositions to better manage employee, user, investor or client performance under challenging information conditions.

ACF posits that the pursuit of optimal results can be best achieved through AI augmentation. AI augmented ACF solutions can be designed with a high degree of adaptability and sensitivity to potential information facets interactions that drive performance. AI augmented ACF solutions can dynamically customize information representations at the individual, team, or organizational levels. These can be integrated with more expansive decision support platforms and information systems applications to provide a wider range of performance improvements.

*3.3. ACF limitations and future research directions*

Unarguably, we live in increasingly complex informational ecosystems, and there is a critical need "*…to study wicked problems in the emerging digital world*" (McKelvey, Tanriverdi, & Yoo, 2016). While ACF helps improve our understanding of ways in which we can mitigate some of the complexity induced by variances in information facets, we identified a few limitations: **First,** as with most experimental studies, there is a need to strengthen external validity with an analysis of data from organizations, exchanges and other real-world environments. While equivocality is easily generalizable, there is a need to validate interaction effects with alternate information representations from multiple domains. **Second,** we note that the experimental study presented in this paper is only one of many potential illustrations of ACF. The ACF framework and propositions are widely generalizable. However, the hypotheses focusing on information facets and information representations are specific to the experiment. Therefore, the scale and scope of their interactions may vary with changes to application domains. **Third,** this study is a foundational articulation of ACF focusing primarily on the core theoretical concepts, and we have been constrained in our ability to vary tasks – it will be useful to add multiple types of tasks in future studies. **Fourth,** the influence of other information facets such as uncertainty and veracity, and dualistic facets such as *overload and dearth* need to be studied and validated for the development of a deeper understanding of ACF (Oh et al., 2013). Most of these limitations pave the way for interesting future research that could develop or qualify the ACF framework.

Additional research is necessary to study how ACF can be deployed using machine learning to improve performance and user satisfaction in a broad range of information systems applications. ACF provides a parsimonious and flexible framework which can be used to study AI driven performance effects with two or more simultaneous information representations. Machine learning methods can help model the effects of complex information facets in scenarios such as stock trading that use multiple display terminals and multiple information representations to reflect the same underlying information. Future research can therefore expand ACF along three dimensions:

1. Extend ACF theory to include additional information facets.
2. Develop ACF models with additional information representations and information artifacts.
3. Apply ACF to multiple practitioner domains, articulate application scenarios for products and services, and measure performance augmentation.

## 4. Conclusion

The AI schema posited by ACF can be generalized to a wide range of future information systems and decision support design solutions, where AI can learn from human performance, and environmental variables, to help us in our pursuit of optimality. The projected trajectory of developments in the big data and AI ecosystems, and machine learning in particular "*have intensified the speed, and our abilities, to create and deploy new knowledge for constructing theories*" (Tremblay, Kohli, & Forsgren, 2021; Abbasi, Sarker, & Chiang, 2016). ACF is a unique forward-looking theory, which aligns well with calls from researchers towards positivism (Kar & Dwivedi, 2020) such as the "*theory in flux*" paradigm (Tremblay et al., 2021): *ACF is deeply rooted in prominent theoretical frameworks, and posits a clear application-oriented design science framework that combines AI with HI in its pursuit of optimal performance.*

Information systems discipline has an established culture of creatively theorizing and modeling the interactions between human behaviors, technology and information from an applied and design science perspective (Tremblay et al., 2021, and in general: MISQ Special Issue: Next Generation IS Theories, March 2021). Keeping in line with this valuable culture, we hope that ACF will provide vital insights to ensure the relevance and applicability of cognitive fit research to emerging big data and AI ecosystems.

**Funding**

We are grateful to Baruch College and the City University of New York, for funding the economic trading experiment.

APC open access support provided by: Rutgers University, William Paterson University & Hofstra University.

**Conflict of interest**

The authors have no known conflict of interest.

**Acknowledgments**

We acknowledge the valuable guidance provided by professors Nanda Kumar, Richard Holowczak and Robert Schwartz, Zicklin School of Business, Baruch College, CUNY.

**Appendix A. Supporting information**

Supplementary data associated with this article can be found in the online version at doi:10.1016/j.ijinfomgt.2022.102505.